# Two Separate Outflows in the Dual Supermassive Black Hole System NGC 6240

F. Müller-Sánchez[1], R. Nevin[1], J. M. Comerford[1], R. I. Davies[2], G. C. Privon[3,4], E. Treister[3]

[1]*Department of Astrophysical and Planetary Sciences, University of Colorado, Boulder, 80309 CO, USA*

[2]*Max-Planck-Institut für Extraterrestrische Physik, Postfach 1312, D-85741 Garching, Germany*

[3]*Instituto de Astrofísica, Facultad de Física and Centro de Astroingeniería, Facultad de Física, Pontificia Universidad Católica de Chile, Casilla 306, Santiago, Chile*

[4]*Department of Astronomy, University of Florida, 211 Bryant Space Sciences Center, Gainesville, 32611 FL, USA*

**Theoretical models and numerical simulations have established a framework of galaxy evolution in which galaxies merge and create dual supermassive black holes (with separations of one to ten kiloparsecs), which eventually sink into the centre of the merger remnant, emit gravitational waves and coalesce. The merger also triggers star formation and supermassive black hole growth, and gas outflows regulate the stellar content[1,2,3]. Although this theoretical picture is supported by recent observations of starburst-driven and supermassive black hole-driven outflows[4,5,6], it remains unclear how these outflows interact with the interstellar medium. Furthermore, the relative contributions of star formation and black hole activity to galactic feedback remain unknown[7,8,9]. Here we report observations of dual outflows**

**in the central region of the prototypical merger NGC 6240. We find a black-hole-driven outflow of [O III] to the northeast and a starburst-driven outflow of Hα to the northwest. The orientations and positions of the outflows allow us to isolate them spatially and study their properties independently. We estimate mass outflow rates of 10 and 75 solar masses per year for the Hα bubble and the [O III] cone, respectively. Their combined mass outflow is comparable to the star formation rate[10], suggesting that negative feedback on star formation is occurring.**

NGC 6240 is an ideal system in which to study the effect of winds on the evolution of a galaxy. Its high star-formation rate (SFR>100 $M_\odot$ yr$^{-1}$ as determined by ultraviolet and infrared luminosity measurements[10]; $M_\odot$, mass of the Sun) and the presence of a dual active galactic nucleus (AGN) with a separation of about 0.7 kpc (bolometric luminosities of 8×10$^{44}$ erg s$^{-1}$ and 2.6×10$^{44}$ erg s$^{-1}$ for the southwestern and northeastern nuclei, respectively[11,12]) ensure ample activity for driving substantial feedback. A nebula with bright optical emission lines is known to be associated with NGC 6240. At scales smaller than 10 kpc, Chandra observations have revealed that the central Hα nebula (the "butterfly nebula") is spatially coincident with the soft X-ray-emitting gas surrounding the dual AGN[11,13,14]. Such good spatial correlation indicates that the hot and warm phases of the interstellar medium in the butterfly nebula are excited via the same mechanism (probably shocks caused by outflowing gas[14,15]). In addition, Hα emission extends to about 90 kpc and shows multiples loops, bubbles and filaments[14,15,16]. Previous studies have suggested that the extended Hα nebula is primarily excited by shock heating induced by a starburst-driven superwind. However, the kinematics of the butterfly nebula has not been studied in detail because high-spatial-resolution observations are needed to separate the two nuclei and identify small-scale kinematic structures.

In this work we use Hubble Space Telescope (HST) imaging and spectroscopic data from the ground-based long-slit Dual Imaging Spectrograph at the Apache Point Observatory (APO/DIS) and the integral-field Spectrograph for Integral Field Observations in the Near-Infrared at the Very Large Telescope (VLT/SINFONI; see Methods) to determine the morphology and kinematics of the low-ionization (Brγ and Hα) and high-ionization gas ([O III] at wavelength λ=5,007 Å; hereafter, [O III]), in the butterfly-shaped inner region of NGC 6240. The HST images reveal the detailed structure of the narrow-line region in NGC 6240 (as traced by the [O III]-emitting gas, which is less affected by star-formation than Hα or Brγ[17,18]) and allow us to ascertain whether the contribution of the dual AGN in the formation of outflowing winds is substantial.

In Fig. 1 we present a three-colour composite image of the central 25″×25″ of NGC 6240, obtained with Wide Field Camera 3 (WFC3) of the HST. Green corresponds to the continuum (filter F621M, with a central wavelength of 6,200 Å), red to Hα (F673N), and blue to [O III] (FQ508N) emission. In the inset of Fig. 1, we summarize the relevant morphological structures. In all WFC3 images there are two emission peaks, which are located at the positions of the two AGNs, whereas most of the continuum emission is in the disk of the merger remnant. Therefore, the extended emission in the F621M image traces the stellar continuum. The prominent dark dust lane running along the northeast-southwest direction is also spatially coincident in the three WFC3 images.

The F673N image reveals filaments and bubbles of Hα emission to both the east and west of the nuclei. The bubbles do not appear to be coincident with the stellar continuum, which suggests that they result from winds, rather than gas associated with tidal debris. Furthermore, the bubbles and filaments to the west of the nuclei are seen mostly in Hα (regions 2 and 4). This emission is probably due to shock ionization from stellar winds (see Methods). On the other side of the nuclei, to the east/southeast of the southwestern nucleus (region 3), knots and filaments are seen in both Hα and [O III], consistent with a scenario in which both the dual AGN and star formation contribute to the ionization of the extraplanar gas. In addition to region 3, the FQ508N image also reveals diffuse [O III] emission in some parts of the galactic disk and an extended extraplanar structure with conical geometry to the northeast of the nuclei (region 1), which is faint to non-existent in Hα. The [O III] gas in region 1 is not spatially correlated with the stellar continuum (Fig. 1) or with the molecular gas (Extended Data Fig. 2), which indicates that this structure is not associated with tidal tails from the merger. These four observations, as well as the similarity between the morphology of the [O III] emission (which extends to a distance of 3.7 kpc to the northeast with an opening angle $\theta_{out} \approx 50°$, see Extended Data Fig. 1) with ionization cones seen in prototypical Seyfert galaxies[17,18], suggest that the gas in region 1 is mostly ionized by the AGN. This interpretation is supported by optical emission-line diagnostics (see Methods). Most of the points in the [O III] cone lie in the Seyfert or LINER (low-ionization nuclear-emission-line region), rather than in the starburst (H II) region, on the Baldwin-Phillips-Terlevich (BPT) diagram[19] (Extended Data Fig. 3) with 20% (3 out of 15) of the spatial elements located in the Seyfert region. In addition, the [O III]/Hβ ratio is considerably enhanced in region 1, reaching a maximum of about 10 (the largest in the butterfly nebula), placing the [O III] cone firmly in the Seyfert region[19,20,21]. This behavior is not observed in any other part of the galaxy and suggests a substantial contribution from AGN photoionization. Because LINERs do not usually show ionization cones of highly-ionized gas[22,23], the [O III] cone is probably produced by the southwestern nucleus, which is also the more powerful of the two AGNs in the system. However, we cannot rule out a contribution from the northeastern nucleus. The location of the [O III] cone (next to the northeastern nucleus) and the fact that the apparent base of the cone encompasses both nuclei, suggest a combined narrow-line region from the two AGNs.

Figure 2 shows the SINFONI velocity map of ionized hydrogen (Brγ) overlaid on the Hα HST image. The velocity maps of the stars and molecular hydrogen ($H_2$) are also shown for comparison. While two rotational components are observed in the stellar kinematics, the $H_2$ velocity map shows one large perturbed rotating disk with a rotation axis that is not aligned with those of the two nuclei. The decoupled $H_2$ disk is probably produced by the tidal forces of the interaction and its sense of rotation follows the orbital history of the merger[24,25]. The rotational velocity of the $H_2$ disk is about 220 km s$^{-1}$. Interestingly, the kinematics of the low-ionization gas is considerably different from both that of the stars and that of $H_2$. Redshifted velocities of 360 km s$^{-1}$ and broad emission lines (velocity dispersion $\sigma \approx 450$ km s$^{-1}$) are observed in the northeastern nucleus at the base of the Hα bubble (Extended Data Fig. 4), indicating an outflow of ionized hydrogen (see Methods).

Position-velocity diagrams of the [O III] emission in region 1 reveal the typical signatures of outflows (Fig. 3): (i) high-velocity (up to 350 km s$^{-1}$) components that cannot be explained by the same gravitational potential that produces H$_2$ velocities of about 220 km s$^{-1}$ in the galaxy disk (Fig. 2), (ii) broad components of [O III] ($\sigma \approx$ 1070 km s$^{-1}$), and (iii) signatures of radial acceleration and deceleration[5,18,26]. These features support the hypothesis of a non-gravitational force accelerating the gas from about 0 km s$^{-1}$ at the centre of the galaxy (between the two nuclei, see Extended Data Fig. 1) to 350 km s$^{-1}$ at a distance $r$=1.8 kpc and subsequently decelerating it to about 190 km s$^{-1}$ at $r$=3.7 kpc (Fig. 3).

The morphology, kinematics, timescale and energetics of the [O III] cone are consistent with energy injection from the AGN (see Methods). In particular, the kinetic power of the outflow is about 2.9 times larger than the estimated injection of energy from the nuclear starburst (assuming SFR=100 M$_\odot$ yr$^{-1}$), requiring a substantial contribution from the AGN. On the other hand, the timescale of the H$\alpha$ bubble (7.4 Myr) and its energetics are consistent with energy injection from a recent episode of star formation that started less than 9 Myr ago. This timescale is inconsistent with the typical AGN flickering cycles[27,28,29], but agrees with the age of the nuclear starburst in NGC 6240[14,25,30].

The AGN-driven outflow carries about 7.5 times more mass ($\dot{M}_{AGN} \approx$ 75 M$_\odot$ yr$^{-1}$) and is about 15 times more powerful (higher kinetic luminosity, $\dot{E}_{AGN} \approx 2 \times 10^{44}$ erg s$^{-1}$) than the outflow in the H$\alpha$ bubble ($\dot{M}_{bubble} \approx$ 10 M$_\odot$ yr$^{-1}$, $\dot{E}_{bubble} \approx 1.3 \times 10^{43}$ erg s$^{-1}$). We note that $\dot{M}_{bubble}$ does not correspond to the total outflow rate due to star formation in the nuclear region ($\dot{M}_{SF}$); regions 3 and 4 also need to be included. Assuming the same properties of the Hα bubble (geometry, kinematics and mass) for regions 3 and 4 (Fig. 1, see also refs [14] and [15]), $\dot{M}_{SF}$ would be about 30 M$_\odot$ yr$^{-1}$. We use the ratio of the mass outflow rate to the SFR to evaluate the influence of negative feedback on the newly formed galaxy disk[6]. A value smaller than 1 indicates that the outflow does not carry enough mass to affect the stellar production considerably. If this ratio is equal to or greater than 1, negative feedback on star formation is occurring. In NGC 6240 the combined effect of $\dot{M}_{AGN}$ and $\dot{M}_{SF}$ is comparable to the SFR[10]. Therefore, we are witnessing the crucial phase in the evolution of mergers of gas-rich galaxies in which suppression of star formation is starting to occur. It is important to note that the starburst-driven outflow alone underestimates the effect of feedback in the galaxy (a similar conclusion is reached for the AGN-driven outflow). Only the combined mass outflow rate can limit the star-formation activity and the growth of the newly formed galaxy after the merger event.

**Acknowledgments**: Some of the data presented in this paper were obtained from the Mikulski Archive for Space Telescopes (MAST). The Space Telescope Science Institute is operated by the Association of Universities for Research in Astronomy, Inc., under NASA contract NAS5-26555. The optical spectroscopic data reported here were obtained at the Apache Point Observatory 3.5m telescope, which is owned and operated by the Astrophysical Research Consortium. F. M-S. acknowledges financial support from NASA HST Grant HST-AR-13260.001. G. C. P. acknowledges support from a FONDECYT Postdoctoral Fellowship (number 3150361) and the University of Florida. E.T. acknowledges support from CONICYT Anillo ACT1101. FONDECYT regular grants 1120061 and 1160999 and Basal-CATA PFB-06/2007.

**Author Contributions:** F.M-S. conceived the project, analysed the data, coordinated the activities and prepared the manuscript. R.N. prepared and reduced the APO/DIS observations, and created the BPT diagrams. F.M-S. and J.M.C. analyzed the *HST* images. R.I.D. reduced the VLT/SINFONI data. E.T. and G.C.P. contributed to the analyses and discussion. All authors discussed the results and implications and commented on the manuscript at all stages.

**Author Information:** Reprints and permissions information are available at www.nature.com/reprints. The authors have no competing financial interests. Correspondence and requests for materials should be addressed to F. Müller-Sánchez (francisco.mullersanchez@colorado.edu).

**Figure 1**. **Three-colour composite image of Hα (red), [O III] (blue), and *V*-band continuum (green) emission in NGC 6240, obtained by the HST.** The major structures identified in the central 25″×25″ of NGC 6240 are labelled. The inset shows the morphological structures identified in the butterfly-shaped region of NGC 6240 using HST narrow-band filters. The northeastern nucleus shows a typical LINER spectrum and is about three times fainter than the southwestern nucleus, which exhibits the characteristics of a heavily obscured Seyfert 2 galaxy[11,12]. The ionized gas in regions 1-4 is extraplanar. The [O III] cone extends to about 4 kpc to the northeast (which is faint in Hα), which indicates an AGN-driven outflow, whereas the Hα bubble to the northwest is indicative of a starburst-driven outflow. All physical units in this paper are based on a concordance, flat *ΛCDM* cosmology with a Hubble constant of 70 km s$^{-1}$ Mpc$^{-1}$. At the redshift of NGC 6240 (*z*=0.0245), 1″ corresponds to about 0.5 kpc.

**Figure 2. VLT/SINFONI maps of NGC 6240.** (**a**) Stellar velocity. (**b**) H$_2$ velocity map. (**c**) Velocity map of Brγ overlaid on the HST image of Hα. In all panels, the contours delineate the *K*-band continuum emission and are spaced at 10% of the peak flux. Although two rotational components are observed in the stellar kinematics, the H$_2$ kinematics shows one perturbed rotational component with a kinematic major axis oriented at a position angle of 22°. The Brγ kinematics of the northeastern nucleus is dominated by non-circular motions. Redshifted velocities of 360 km s$^{-1}$ are observed at the base of the Hα bubble. In all maps north is up and east is to the left. RA, right ascension; dec., declination. The colour bar indicates line-of-sight velocity (V$_{LOS}$) in units of kilometres per second.

**Figure 3**. **Kinematics of [O III]. a, b,** Segments of the two-dimensional long-slit spectra of NGC 6240 centred at the rest wavelength of [O III]. The colour scale represents flux density normalized to the peak. Cool colors (green and blue) correspond to background emission (<10% of the peak of emission). Warm colours (yellow to red) correspond to sizeable flux density values (>10% of the peak of emission). **c, d,** Position-velocity diagrams of [O III] and $H_2$ emission, where $v$ is the velocity and $\sigma$ is the velocity dispersion. The galaxy was observed at two position angles, 22° and 56° (see also Extended Data Fig. 1). Positive values of angular distance (vertical axis in **a** and **b** and horizontal axis in **c** and **d**) correspond to the direction north from the centre of the galaxy, at the marked position angle. The number of spatial elements extracted from our long-slit observations (**a, b**) is 27 at $PA_1 = 22°$ (**c**) and 34 at $PA_2 = 56°$ (**d**). There are 15 spatial elements inside the [O III] cone (between 0" < $r$ < 7" in **d**). We extracted the velocity and dispersion values for $H_2$ at 7 different spatial positions along imaginary APO/DIS long slits oriented at 22° and 56° in the SINFONI data (Fig. 2).

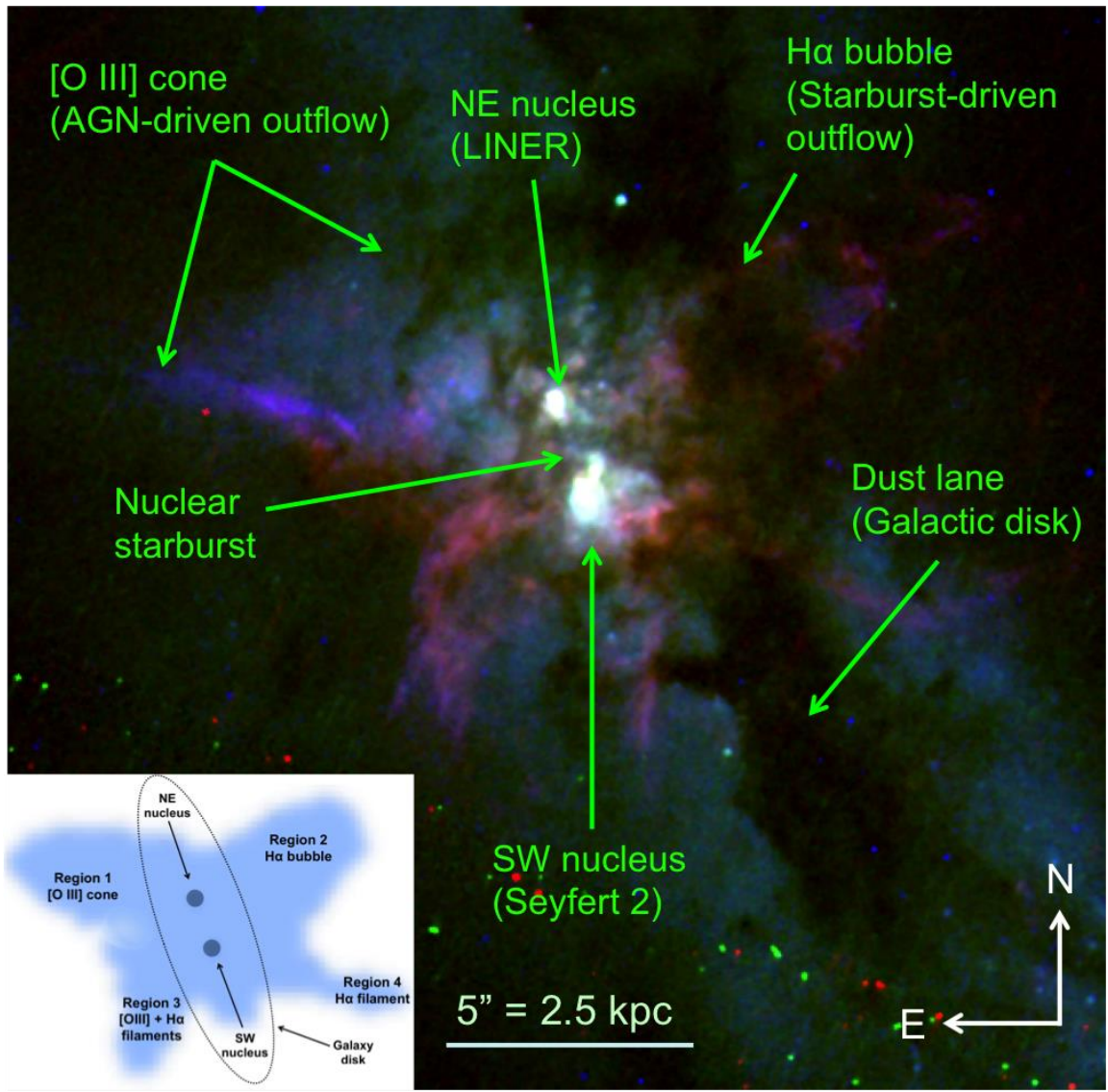

Figure 1.

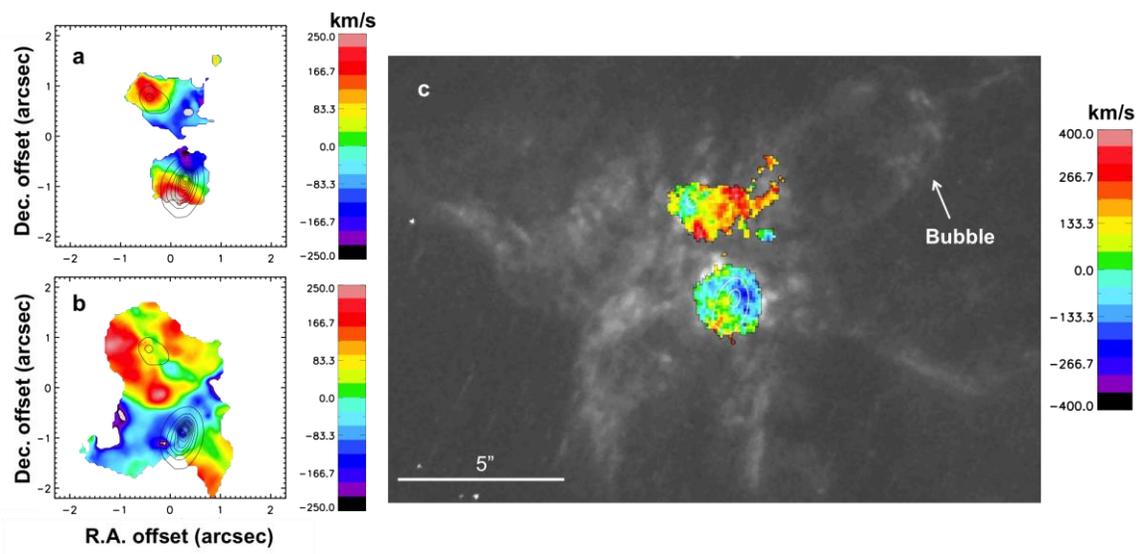

Figure 2.

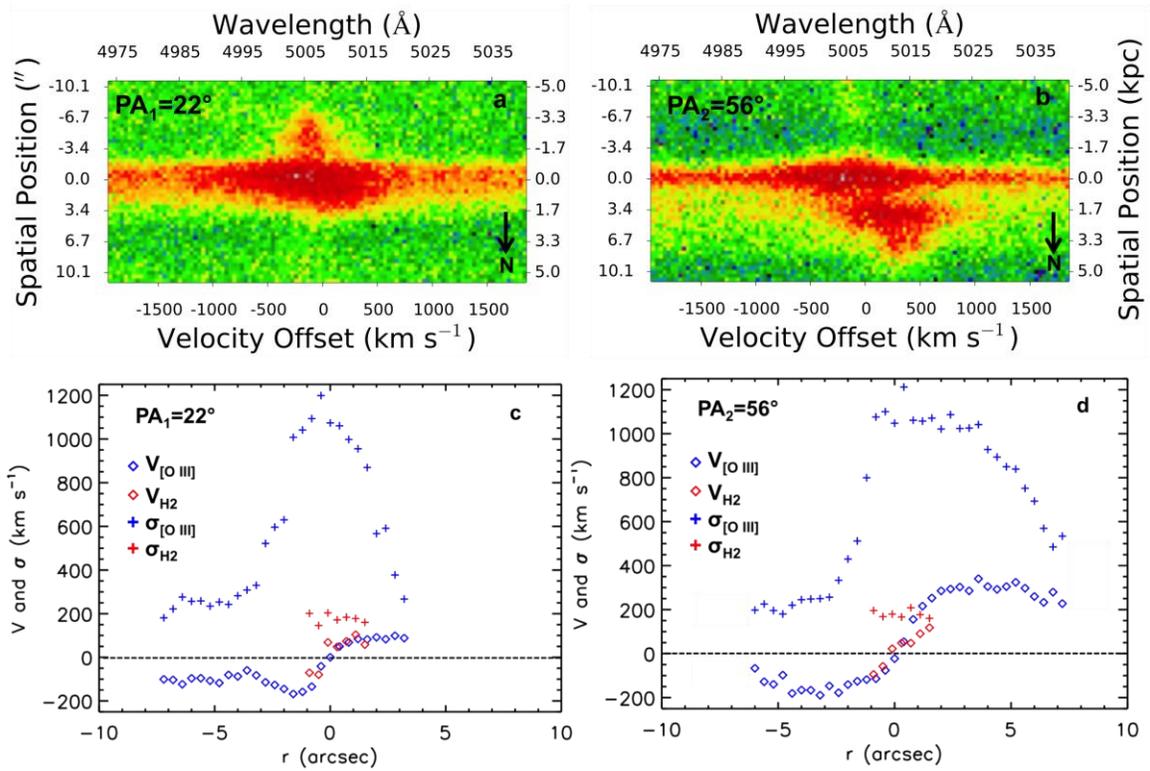

Figure 3.

**Methods**

**HST imaging.** NGC 6240 was observed with the WFC3 on 2 August 2012 (GO-12552; principal investigator, L. Kewley). The HST/WFC3 data provide high-resolution (0.0396″ per pixel) optical imaging in five narrow-band and two medium-band filters (WFC3-UVIS channel). In this work, we selected the filters FQ508N and F673N as tracers of the high-ionization ([O III]) and low-ionization (Hα) gas, respectively. At the redshift of NGC 6240 ($z=0.0245$), these two filters entirely cover the [O III] and Hα emission. Blue (F467M) and red (F621M) continuum images were also used to trace the emission of the stars in the system. We used continuum-subtracted [O III] and Hα images for the analysis. To obtain an image of the ionized gas emission ([O III] and Hα), we aligned the off-band image with the on-band one, scaled it to fit the on-band image at radii where the ionized gas flux is negligible, and subtracted it from the on-band image.

**Near-infrared adaptive-optics-assisted integral-field spectroscopy.** The integral-field data of NGC 6240 used in this work are VLT/SINFONI observations[31,32], obtained on the night of 20 August 2007 (programme 079.B-0576). Details on the observations and data reduction have been described in a previous publication[30]. The final datacube has a spatial resolution of 0.097″×0.162″ full-width at half-maximum (using a pixel scale of 0.05″ per pixel and a field of view of 3.6″×4.0″) and a spectral resolution of about 90 km s$^{-1}$ full-width at half-maximum in the *K*-band. We derived the two-dimensional properties (flux and velocity maps) of the gas and the stars using the IDL code LINEFIT[33], which estimates the uncertainties in the Gaussian fits using Monte Carlo techniques. This method fits the emission lines (absorption features) by convolving a Gaussian with a spectrally unresolved template profile (a sky line for emission lines, a template stellar spectrum for stellar absorption features) to the continuum-subtracted spectral profile at each spatial pixel in the datacube. The velocities are measured relative to the systemic redshift of the galaxy (z=0.0245), taking into account in the stellar kinematics the 50 km s$^{-1}$ redshift of the northeastern nucleus with respect to the southwestern nucleus[25,30]. Several realizations (usually 100) of the data ae generated by adding random noise to the flux at each pixel; these are fitted using the same procedure as above. This method allowed us to obtain uncertainties for the kinematic maps in the range of 30-40 km s$^{-1}$. We note that the near-infrared spectrum of NGC 6240 does not show any strong high-ionization lines from the narrow-line region (the coronal lines [Si VI] and [Ca VIII] are not present in NGC 6240[30]), and therefore a new optical spectrum was acquired.

**Optical long-slit spectroscopy.** The long-slit spectroscopic observations of NGC 6240 were obtained on 30 June 2016 using the DIS with a 1.5″×6' slit at the APO. We used a grating with 1200 lines per millimetre to obtain a spatial resolution of 0.62 Å per pixel. The spatial scales were 0.4″ per pixel and 0.42″ per pixel in the red and blue channels, respectively. We observed with two slit positions for twenty minutes at each position. Slit position 1 (PA$_1$ hereafter) was oriented at 22° (the position angle is measured counter-clockwise, from north, 0°, to east, 90°), whereas slit position 2 (PA$_2$ hereafter) was oriented at 56° (see Extended Data Fig. 1). The emission lines were modelled as Gaussians. For each row in the spatial direction, we fitted a single Gaussian combined with a straight line with a given slope (representing the continuum emission) to each emission line. The uncertainties were estimated using Monte Carlo techniques. The method involves adding noise to the galaxy spectra and refitting the result (about 500 times) to empirically determine the standard deviation of each fitted parameter. Using this method we obtained uncertainties of 70–160 km s$^{-1}$ for the velocity measurements and flux errors of up to 25%.

**Emission-line diagnostics.** In Extended Data Fig. 3 we show the BPT[19] emission-line diagnostic diagrams for all the spatial positions along the two slits where the relevant emission lines - Hβ, [O III], Hα and [N II] at λ=6,584 Å (hereafter [N II]) - have a signal-to-noise ratio greater than 3. To construct the BPT diagrams for each position angle, we measure emission-line fluxes at all spatial positions of the galaxy (for Hβ, [O III], Hα, and [N II]). We first determine the spatial center of the galaxy from the galaxy continuum (Fig. 3) and then fit a single Gaussian for Hβ and [O III]. We fit three Gaussians for the Hα-[N II] complex, where we require the velocity dispersions of the [N II] lines to be identical and their flux ratio to be 1:3.

Most of the emission in the galaxy disk ($PA_1=22°$) is consistent with starburst/LINER excitation. Two data points are located in the border between the star-forming and the Seyfert regions of the diagram, but these could be artifacts caused by a low signal-to-noise ratio (weak [O III] and Hβ emission is observed in the north part of the galaxy disk; Figs. 1 and 3). Along the direction of $PA_2=56°$, the situation is different. At distances between -6″ and -1″ from the nucleus (distances are negative to the south), the emission is located in the star-forming part of the BPT diagram. Between -1″ and 1″ (the centre of the galaxy), the [N II]/Hα ratio increases from about 0.3 to about 5, shifting data points towards the LINER region of the diagram. Finally, inside the [O III] cone, at distances between 1″ and

6″, the [O III]/Hβ ratio increases from about 2 to about 10, shifting data points towards the Seyfert region of the diagram. However, because the [N II]/Hα ratios are also high (>2), some points remain in the LINER region. High [N II]/Hα values (>1.5) usually correspond to LINER-like excitation from shocks[34,35,36]. Therefore, the region occupied by the [O III] cone in the BPT diagram suggests the presence of strong shocks and AGN photoionization.

From our SINFONI data, we also obtained a map of the $H_2$/Brγ ratio (Extended Data Fig. 5). This map is usually used as a tracer of shocks in the near-infrared. If $H_2$ is excited by ultraviolet photons from stars, this ratio should be small since the photons from those stars should also produce substantial Brγ emission (star-forming regions have $H_2$/Brγ ratios[37,38,39] <0.6). A high $H_2$/Brγ value (>1) would indicate shocks rather than star-formation[36,40]. As can be seen in Extended Data Fig. 5, our data show very high $H_2$/Brγ ratios (between 3-48) with a maximum of 48 located between the two nuclei. Therefore, the bright $H_2$ emission in the central region of NGC 6240 is produced by shocks. The large $H_2$/Brγ ratio of 48 (the largest value in the local Universe) is probably produced[41] by the collision of the interstellar media associated with the two progenitor galaxies. In addition, the combination of moderately high $H_2$/Brγ ratios (between 1-15) with high-velocity dispersion values ($\sigma$>250 km s$^{-1}$) is usually associated with outflows[35,36,41]. These two characteristics are observed at the inner edge (or base) of the Hα-bubble (Extended Data Figs. 4 and 5), suggesting the presence of an outflowing wind in this region.

Our results confirm previous studies of the optical and near-infrared emission-line ratios in NGC 6240, which suggest that the emission in the central 3.0″×3.0″ region of the galaxy is dominated by shocks, showing a typical shock-excited LINER spectrum. The near-infrared data imply the presence of two types of shocks: (i) strong shocks, with high $H_2$/Brγ ratios (higher than 15), between the nuclei, caused by the collision of the interstellar media of the merging galaxies and (ii) shocks with slightly lower $H_2$/Brγ ratios (3-15) and higher velocity dispersions (up to about 450 km s$^{-1}$), produced by outflowing winds. The inner edge of the Hα bubble is consistent with the latter type of shocks. Our APO/DIS observations provide information on the excitation mechanisms of the ionized gas in the [O III] cone. Our BPT diagrams suggest the presence of shocks and AGN photoionization in this region. Finally, our long-slit data indicate that the nebular emission in region 4 is dominated by star formation.

**Evidence for outflows in the Hα bubble**. Several pieces of evidence suggest that the kinematics of the Hα bubble is dominated by outflows. First, the ionized gas (Brγ) in the northeastern nucleus exhibits line-of-sight velocities of 360 km s$^{-1}$, which are too high to be explained by the same gravitational potential that is producing rotational velocities of the stars of 200 km s$^{-1}$ in this nucleus (Fig. 2). Furthermore, the velocity map of Brγ is very different to that of the stars. The kinematic major axis of the stars has a position angle of about 37°, which is consistent with the photometric major axis of this nucleus (obtained from adaptive-optics images[30,42]). The kinematic major axis of Brγ extends in the east-west direction with a particularly fast component to the northwest (360 km s$^{-1}$), which is spatially coincident with the Hα bubble seen in HST images. A similar result is obtained when comparing the Brγ velocity map with that of $H_2$. The molecular gas disk has a rotation axis that is not aligned with that of either nucleus[24,25]. The $H_2$ disk has a position angle of 22° and exhibits redshifted velocities of 220 km s$^{-1}$ in the northeastern nucleus. The velocity map of Brγ shows both, redshifted and blueshifted velocities in this nucleus. In addition, the maximum Brγ velocity of 360 km s$^{-1}$ is inconsistent with the maximum rotational velocity of the molecular gas in the disk of the advanced merger (220 km s$^{-1}$). Finally, the gas at the base of the Hα bubble has a very high velocity dispersion (450 km s$^{-1}$), which can be explained only by outflows at these scales[4,5,35,43]. All these results support strongly the premise of a non-gravitational force that is accelerating the gas in the northeastern nucleus to a maximum line-of-sight velocity of about 360 km s$^{-1}$ at $r$=3.3″ (1.6 kpc) with a maximum velocity dispersion of about 450 km s$^{-1}$. These results are broadly consistent with those of a previous study[15], which found a maximum line-of-sight velocity of about 400 km s$^{-1}$ for the region covered by the Hα bubble in seeing-limited long-slit observations of NGC 6240.

**Evidence for outflows in the [O III] cone.** Two structural features suggest strongly the presence of outflows in the [O III] cone: (i) the location of the [O III] cone is outside the plane of rotation of the disk of the merger, in a region that is not associated with tidal structures from the galaxy merger (see Extended Data Fig. 2), and (ii) its morphology is typical of outflows seen in prototypical starburst and Seyfert galaxies[5,6,8,44]. Conical morphologies are not expected for inflows, which are typically radial streamers of gas[45,46]. These two characteristics basically rule out rotation or inflows as possible kinematic components of the [O III] emission in region 1.

The kinematics of the [O III] cone indicates the presence of outflows. Figure 3 shows the two-dimensional APO/DIS spectra of NGC 6240 and the kinematics (position-velocity diagrams) of [O III], extracted at the two position angles (PA$_1$ and PA$_2$) indicated in Extended Data Fig. 1. For comparison, we also extracted the kinematics of H$_2$ along the directions of PA$_1$ and PA$_2$, matching the width of the optical slits. At both position angles, the spectra exhibit a broad component of [O III] ($\sigma$=1,220±140 km s$^{-1}$) in the central 1.5″, with a line-of-sight velocity consistent with the systemic velocity of the galaxy. These extremely broad lines suggest the existence of an outflow that originates from the central 1.5″ of the galaxy (the width of the DIS long-slit). In addition, one clear trend emerges: the gas is more kinematically disturbed along the direction of the morphologically inferred outflow (PA$_2$=56°) than along the disk of the merger. At PA$_1$=22°, the [O III] emission is blueshifted by about 180 km s$^{-1}$ in the south, redshifted by about 100 km s$^{-1}$ in the north, and the emission lines are narrow ($\sigma$ < 250 km s$^{-1}$), consistent with the curves of H$_2$.

By contrast, at PA$_2$=56°, the redshifted broad emission lines to the north provide support for an outflow-dominated kinematic component. As can be seen in Fig. 3, the [O III] velocity curve deviates considerably from that of H$_2$ at distances $r$>1″. The H$_2$ velocity reaches a maximum of 140 km s$^{-1}$ at $r$=2″. At this distance, the [O III] velocity is about 250 km s$^{-1}$, and it continues to increase with distance, reaching a maximum velocity of $v_{max}$=350±30 km s$^{-1}$ at $r$=3.6″. This velocity is too high to be explained by the gravitational potential of the H$_2$ disk. The maximum dispersion of [O III] is 1,070±110 km s$^{-1}$ at $r$=2.1″. We adopted this value as the representative velocity dispersion of the nebula because the velocity 1,220 km s$^{-1}$ at $r$≈0″ might be affected by other random motions caused by the merger process (see Extended Data Fig. 5). The outflow component then begins to decelerate outside this maximum-velocity region, reaching about 190 km s$^{-1}$ at $r$=7.4″, which is an observational signature of an outflow encountering drag forces from the interstellar medium[5,26]. In the south region, along the direction of PA$_2$, the [O III] emission is faint and narrow, with line-of-sight velocities consistent with the rotation of the disk of the advanced merger (Fig. 3). Finally, the broad kinematic components in combination with the high [N II]/Hα ratios (characteristic of shock ionization; see Extended Data Fig. 3), provide strong evidence for the existence of outflowing gas in the [O III] cone[35,36].

**Estimation of mass outflow rates.** The amount of feedback, in terms of outflowing mass entrained in the [O III] cone and Hα bubble, can be estimated using the morphological parameters derived from the HST images and the velocities measured in our long-slit and integral-field data. For the [O III] cone, we estimate the mass outflow rate ($\dot{M}_{cone}$) using a method described in an earlier publication[5] and by assuming a gas density of $n_e$=50 cm$^{-3}$ and a filling factor of $f$=0.01, which are typical of the narrow-line region at $r$≈1.5 kpc[14,15,47]. The [O III] cone has an opening angle of 50±3° (Extended Data Fig. 1), and the projected distance at which the outflow reaches $v_{max}$=350 km s$^{-1}$ is $r_t$ = 1.8 kpc. At $r$>$r_t$ the deceleration phase starts (see Fig. 3d). We have derived the characteristic outflow speed as $v_{out}$=$\sqrt{v_{max}^2 + \sigma_{mean}^2}$/sin $i$, where $i$ is the inclination and $\sigma_{mean}$=505±120 km s$^{-1}$ is the average velocity dispersion of the gas inside the [O III] cone (Fig. 3). This term takes into account the spread of velocities along the line of sight and additional turbulence that may be substantial. Although our geometric model of the [O III] cone does not constrain $i$, it excludes values smaller than 25° (which would imply that the face of the cone that is closer to us is blueshifted) and greater than 65° (a nearly face-on view to the cone). We assume sin $i$ ≈ cos $i$ ≈ 0.71 to account for the unknown inclination of the outflowing gas (this correction is also applied to the distance at which the outflow reaches its maximum velocity $r_t$). The resulting mass outflow rate is $\dot{M}_{cone}$=75 M$_\odot$ yr$^{-1}$ with an uncertainty of a factor of three. We were able to mitigate the uncertainties in the geometrical parameters and kinematics of the [O III] cone thanks to the high-resolution HST images and our detailed curves of line-of-sight velocity and velocity dispersion as a function of position. The uncertainty of a factor of three in $\dot{M}_{cone}$ comes from the assumed value of density, which for NGC 6240 and other galaxies with high infrared luminosities is estimated to be in the range[15] 20-150 cm$^{-3}$, with a typical value of 50 cm$^{-3}$ at $r$=2 kpc (in other words, log $n_e$=1.7±0.47). We point out that the obtained value of $\dot{M}_{cone}$ is probably a conservative estimate for the mass outflow rate because we may have underestimated the outflow covering factor (we considered only the region inside the conical structure of [O III] emission but the outflow might be covering a larger volume) and the maximum outflow velocity (up to 1,070 km s$^{-1}$ for the high-dispersion [O III] clouds).

For the Hα bubble we calculated the outflow rate as[27] $\dot{M}_{bubble}=M_{H\alpha}v_{out}/r$ because the mass of ionized hydrogen is known. A mass of 1.4×10$^8$ M$_\odot$ has been estimated[14] for the butterfly-shaped nebula, assuming a spherically-symmetric structure. The Hα bubble covers approximately one quadrant of the butterfly nebula (Fig. 1). Therefore, the mass of ionized hydrogen in this region is about 3.5×10$^7$ M$_\odot$. The maximum velocity of the gas is 360 km s$^{-1}$ at the inner edge of the bubble (Fig. 2) and $\sigma_{mean}$=260±40 km s$^{-1}$ (Extended Data Fig. 4). The Hα bubble appears to be perpendicular to both the galaxy disk and the northeastern nucleus, which has an inclination[25,30] of about 45°. Therefore, we again adopt sin $i$ ≈ cos $i$ ≈ 0.71 as a de-projection factor. Assuming the bubble is expanding at a constant velocity up to a radius of 1.6 kpc, the resulting $\dot{M}_{\text{bubble}}$ is 10 $M_\odot$ yr$^{-1}$. The largest uncertainty in the calculation of $\dot{M}_{\text{bubble}}$ comes from the mass of ionized gas, which has been estimated[14] as 20%. However, we are probably underestimating the mass of Hα in the Hα bubble, which can be up to three times larger (taking into account all the hydrogen gas that is being ionized by star formation in regions 3 and 4).

**Driving mechanism of the outflows.** Here we compare the morphologies, velocities, timescales and energetics of the [O III] cone and the Hα bubble to identify the primary driver of the outflow in each region. We calculate the dynamical time of the outflows as $t_{\text{dyn}}=D/v_{\text{out}}$, where $D$ is the size of the [O III] cone or Hα bubble, and obtain 7.4±1.4 Myr for the Hα bubble and 3.9±1.2 Myr for the [O III] cone. The timescale of the Hα bubble is inconsistent with the typical timescale of the active phase of an AGN (the AGN can flicker on and off, showing a variability of 1-3 orders of magnitude, with a timescale of 0.1 Myr to a few million years owing to stochastic accretion at small scales[27,28,29]), but agrees with the age of the most recent starburst in NGC 6240 (6-9 Myr)[14,25,30]. By contrast, the timescale of the [O III] cone is similar to those of typical AGN flickering cycles[29], suggesting the presence of an AGN-driven outflow in region 1 of the butterfly nebula.

Next, we estimate the amount of energy injection required to power the outflows (kinetic power and luminosity). We use two methods[43]: the first method[5] estimates the kinetic power of the outflow as $\dot{E}=\dot{M}v_{\text{out}}^2/2$. The second method treats the outflows as analogous to supernova remnants, but with continuous energy injection, and estimates the energy injection rate required to expand the outflows into a low-density medium. Using the first method we obtain kinetic powers of $1.2\times10^{42}$ erg s$^{-1}$ for the Hα bubble and $1.9\times10^{43}$ erg s$^{-1}$ for the [O III] cone. As mentioned earlier, these are conservative values (probably lower limits) for the kinetic power of the outflows because we are probably underestimating the mass outflow rates. For the second method, we used equation (2) in ref. [43] with an ambient density of an uniform low-density medium $n_0$=0.5 cm$^{-3}$ and a covering factor of 1 (these values probably represent upper limits for these parameters). We found an energy injection rate of $1.5\times10^{44}$ erg s$^{-1}$ and $2\times10^{45}$ erg s$^{-1}$ for the Hα bubble and the [O III] cone, respectively. We adopted a fiducial range between $1.2\times10^{42}$ and $1.5\times10^{44}$ erg s$^{-1}$ for $\dot{E}_{bubble}$ and between $2\times10^{43}$ and $2\times10^{45}$ erg s$^{-1}$ for $\dot{E}_{AGN}$, using single fiducial values (the average of the lower and upper limits of $\dot{E}$ in logarithmic scale) of $1.3\times10^{43}$ erg s$^{-1}$ and $2\times10^{44}$ erg s$^{-1}$ for $\dot{E}_{bubble}$ and $\dot{E}_{AGN}$, respectively. The results of the two approaches indicate that the outflow in the [O III] cone is about 15 times more powerful than the outflow in the Hα bubble, which reflects the fact that the former is faster and more extended than the latter.

We can estimate the amount of mechanical energy returned from the nuclear starburst as[4] $\dot{E}_{mech}$=$7\times10^{41}$ (SFR/M$_\odot$ yr$^{-1}$) erg s$^{-1}$. Assuming an SFR of 100 M$_\odot$ yr$^{-1}$ in the central region of the galaxy, the total injection of energy from the stars is $7\times10^{43}$ erg s$^{-1}$. Thus, star-formation is consistent with powering the outflow in the Hα bubble for the standard value of $\dot{E}_{bubble}$ ($1.3\times10^{43}$ erg s$^{-1}$). In addition, energy injection from the stars could power the outflow in the [O III] cone at the lower limit of the energy range. However, we consider this unlikely, because this energy injection is below the fiducial value of kinetic power required to drive the outflow in the [O III] cone, $2\times10^{44}$ erg s$^{-1}$. The bolometric luminosity of the dual AGN estimated[12] from NuSTAR hard X-ray data is $1.1^{+0.5}_{-0.3}\times10^{45}$ erg s$^{-1}$, which is consistent with the value obtained[13] by fitting the spectral energy distribution (about $2\times10^{45}$ erg s$^{-1}$). Our upper limit on the kinetic power of the [O III] cone is consistent with the bolometric luminosity of the dual AGN. We therefore conclude that the dual AGN is energetically capable of powering the outflow in the [O III] cone without help of the starburst and that the mechanical energy injection rate from star formation is not powerful enough to accelerate the gas in this region.

In general, it is difficult to identify the driving mechanism of outflowing bubbles. In the case of the Hα bubble, four pieces of evidence suggest that the outflow is driven by star formation. First, the nuclear starburst is energetically capable of driving the outflow without the need to invoke energy from an AGN. Second, the dynamical time of the outflow (about 7.4 Myr) is consistent with the age of the nuclear starburst in NGC 6240 (about 6–9 Myr[14,25,30]). Third, from a morphological point of view, a wind perpendicular to a nuclear disk is consistent with the structures of starburst-driven winds, in contrast to AGN-driven outflows, which have random orientations with respect to the galaxy disk[35]. In NGC 6240, the position angle of the major axis of the Hα bubble (about 110°) is almost perpendicular to the disk of the advanced merger (22°). On the other hand, the [O III] cone has a position angle of 56°, randomly oriented with respect to the galaxy disk. Finally,

AGN-driven outflows usually have higher velocities than starburst-driven outflows[4,35]. The [O III] cone exhibits gas clouds with velocity dispersion values of 1,070 km s$^{-1}$, whereas the maximum dispersion of the Brγ emission is about 450 km s$^{-1}$ (see also ref. [15]).

**Data Availability:** The data plotted in the figures and that support other findings of this study are available from the corresponding author upon reasonable request. The SINFONI data used in this paper (programme 079.B-0576) can be obtained from the ESO Science Archive Facility. The HST images (GO-12552) are available at MAST.

**Extended Data Figure 1. Contour image of [O III] emission in NGC 6240.** The blue curves show linear contours for the HST/F502N observations. The contours are set at 7.5%, 15%, 30%, 45%, 60%, 75% and 90% of the peak of emission. The extended [O III] emission is traced by the contours representing 7.5% of the peak of emission. The other contours (15-90% of the peak of emission) are located mostly around the two nuclei. A geometric model of the [O III] cone is shown in light blue. The model was created using the software Shape[48]. We constrained the model (size and opening angle) to follow the outer contours (7.5% of the peak of emission) of the wedge-shaped structure in region 1. Interestingly, a regular cone (a cone with a sharp apex) does not provide a good fit to the wedge-shaped structure. The best fit is obtained for a truncated cone. If we had used a regular cone, the apex would be located exactly at the position of the southwestern nucleus. This is consistent with our interpretation that the [O III] cone is probably produced by the two AGNs, with a larger contribution from the southwestern nucleus. For the [O III] cone, we obtained a size of 3.7±0.2 kpc and an opening angle of 50.2±3.1°. The red-shaded rectangles indicate the spatial coverage of the DIS long-slits. $PA_1=22°$ is oriented along the major axis of the galaxy disk, and $PA_2=56°$ covers the region where the [O III] cone is observed (region 1). Both slits

were centred between the nuclei. The dashed rectangle represents the SINFONI field of view. North is up and east is to the left.

**Extended Data Figure 2. Comparison of the morphologies of [O III] and $H_2$.** An image of $H_2$ emission obtained with adaptive optics and the near-infrared camera 2 (NIRC2)[42] of the Keck Observatory is superimposed on the [O III] contours from Extended Data Fig. 1. Black represents fluxes <0.011 of the peak of emission. The absence of molecular gas at the locations of the [O III] cone and the Hα bubble indicates that these two structures are located in regions that are not greatly influenced by the merger process. By contrast, the majority of perturbations caused by the merger activity are seen in the central region between the nuclei, and as gas streamers in the regions east and southwest of the southwestern nucleus (regions 3 and 4 in our analysis; see also Fig. 2b).

**Extended Data Figure 3. Optical emission-line diagnostic diagrams.** The galaxy was observed at two position angles, $PA_1=22°$ and $PA_2=56°$ (see Extended Data Fig. 1). The positive values of angular distance (green to red in the colour bar) correspond to the direction north of the centre of the galaxy, at that position angle. Negative angular distance values (green to blue in the colour bar) correspond to the direction south of the centre of the galaxy, at that position angle. The BPT diagram is usually divided into three regions: AGN (or Seyfert), LINER (or LIER, low-ionization emission-line region; see also ref. [50]) and H II (or starburst region). In both panels, we plot the extreme starburst diagnostic line[21] (curved dashed line) and the LIER/LINER

diagnostic line[49] (straight dashed line). Hβ emission was detected with a signal-to-noise ratio higher than 3 in 16 spatial elements at $PA_1 = 22°$ (from $r = -6''$ to $r = 2''$) and in 26 spatial elements at $PA_2 = 56°$ (from $r = -5''$ to $r = 6''$). There are 15 spatial elements inside the [O III] cone (Fig. 3). The error bars correspond to the uncertainties of the flux ratios (one standard deviation) and were calculated via standard error propagation for the flux of each emission line.

**Extended Data Figure 4. Map of Brγ velocity dispersion.** The contours delineate the Brγ flux distribution and are set at 15%, 30%, 45%, 60%, 75% and 90% of the peak of emission. The dashed rectangle delimits the base of the Hα bubble. Regions in white correspond to pixels where the Brγ flux is less than 5% of the peak of emission and thus were masked out. North is up and east is to the left. The colour bar indicates the range of velocity dispersion values observed in units of kilometres per second.

**Extended Data Figure 5. Map of $H_2$/Brγ flux ratio.** The contours delineate the Brγ flux distribution and are set at 15%, 30%, 45%, 60%, 75% and 90% of the peak of emission. The dashed rectangle delimits the base of the Hα bubble. Regions in white correspond to pixels where the Brγ flux is less than 5% of the peak of emission and thus were masked out. North is up and east is to the left. The colour bar indicates the range of ratios observed.

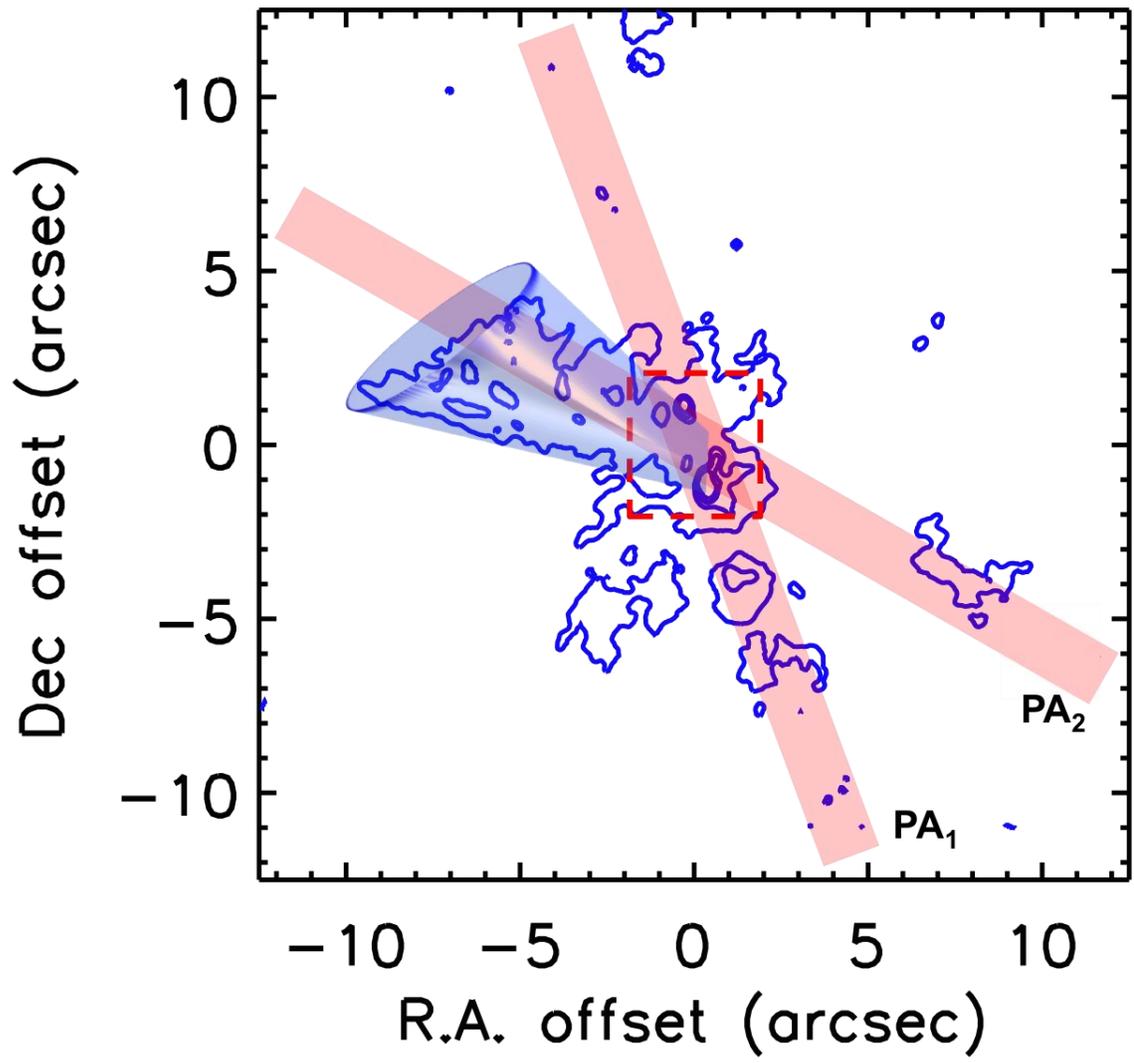

Extended Data Figure 1.

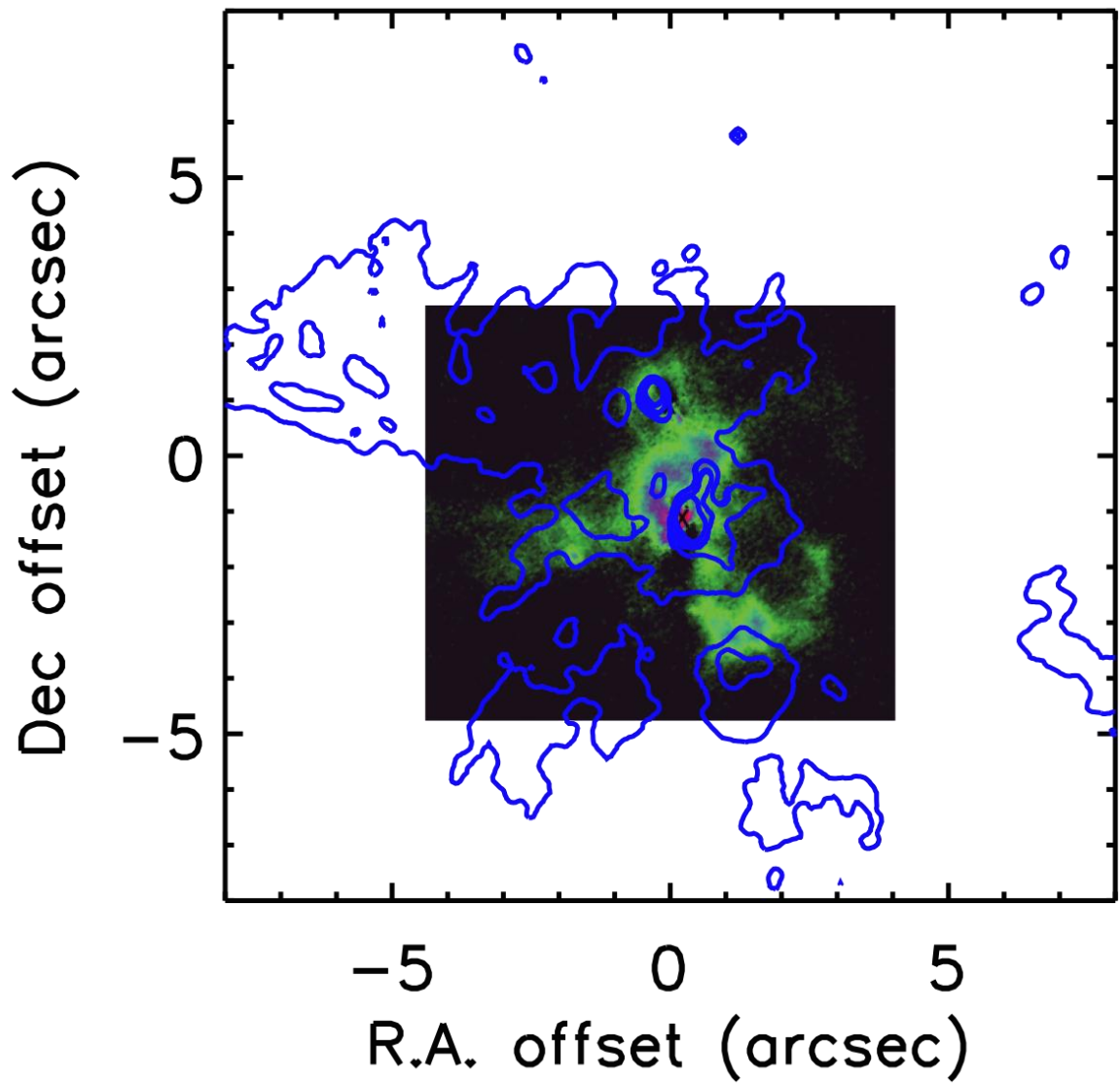

Extended Data Figure 2.

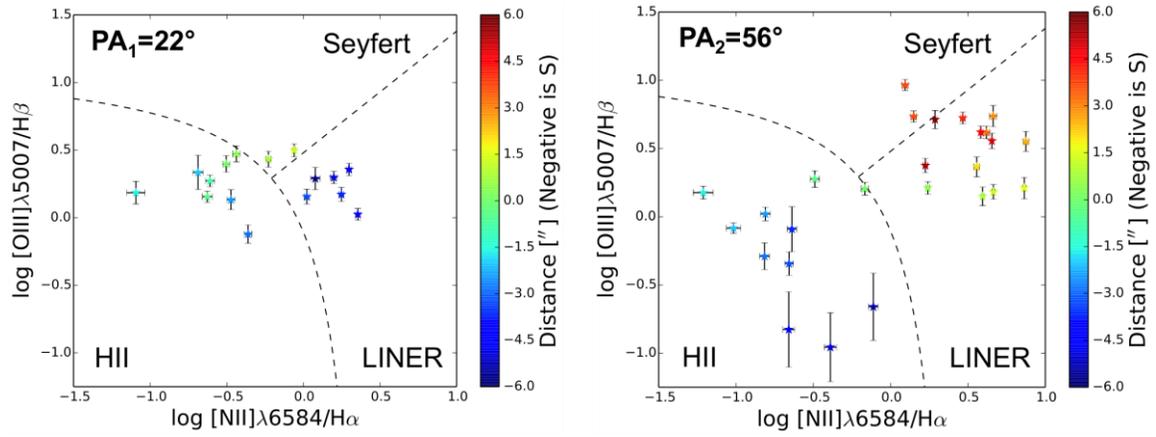

Extended Data Figure 3.

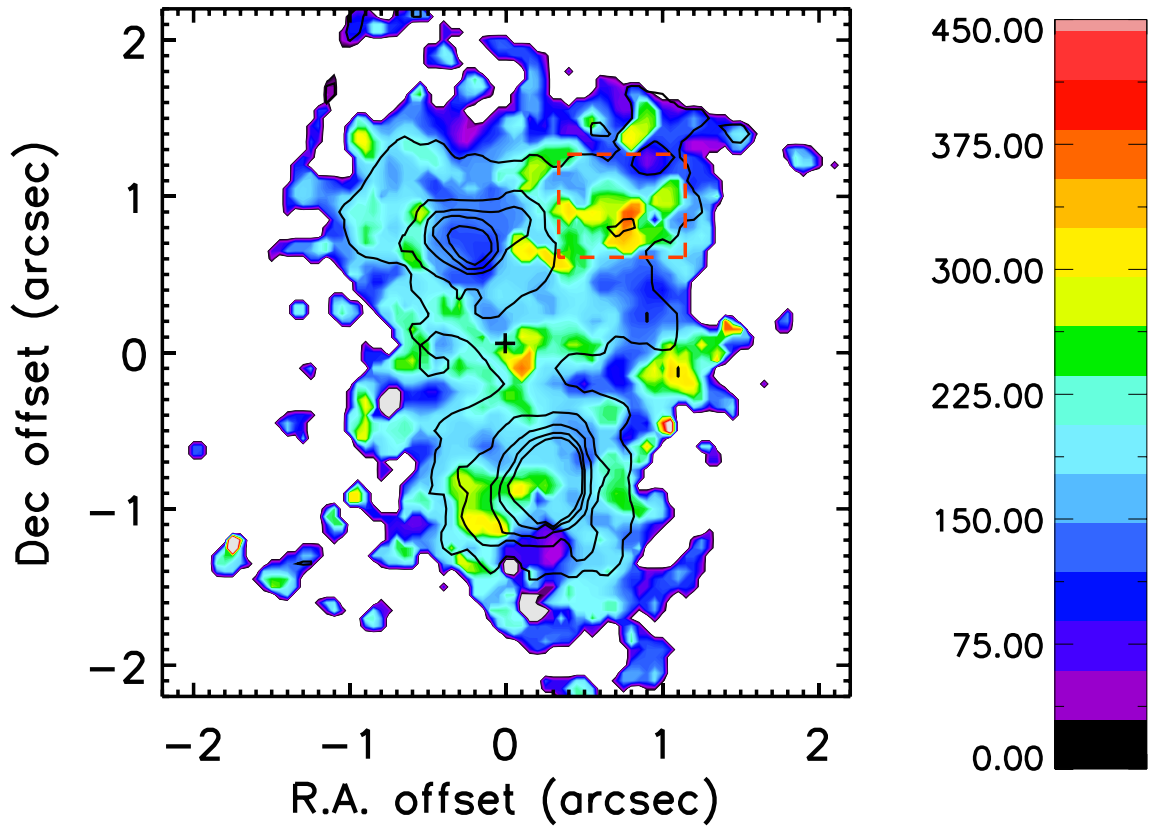

Extended Data Figure 4.

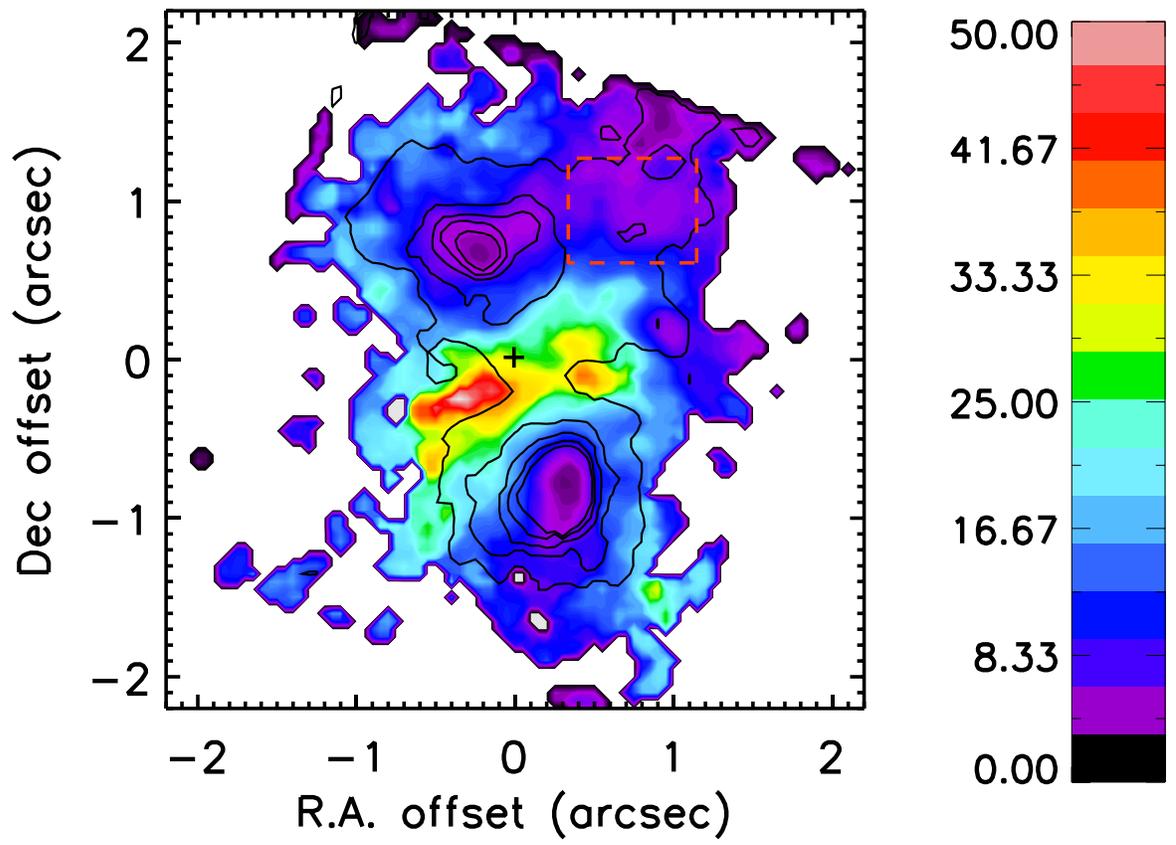

Extended Data Figure 5.